\documentclass{JAC2003}

\usepackage{graphicx}
\usepackage{booktabs}

\begin{document}
\title{
APOCHROMATIC TWISS PARAMETERS OF DRIFT-QUADRUPOLE SYSTEMS WITH SYMMETRIES
\vspace{-0.5cm}}
\author{V.Balandin\thanks{vladimir.balandin@desy.de}, 
R.Brinkmann, W.Decking, N.Golubeva \\
DESY, Hamburg, Germany}

\maketitle

\begin{abstract}
In this article we continue the study of the 
apochromatic Twiss parameters of straight drift-quadrupole systems
with special attention given to the properties of these parameters
for beamlines with symmetries.  
\end{abstract}

\vspace{-0.3cm}
\section{INTRODUCTION}

\vspace{-0.1cm}
A straight drift-quadrupole system can not be designed in such a way that  
a particle transport through it will not depend on the difference 
in particle energies and this dependence can not be removed even 
in first order with respect to the energy deviations.
Nevertheless, the situation will change if instead of 
comparing the dynamics of individual particles  
one will compare the results of tracking of
monoenergetic particle ensembles through the system or
will look at chromatic distortions 
of the betatron functions appearing after their transport through the system.
From this point of view, as it was proven in ~\cite{ApIPAC10},
for every drift-quadrupole system which is not a pure drift space
there exists an unique set of Twiss parameters 
(apochromatic Twiss parameters), which will be transported through
that system without first order chromatic 
distortions\footnote{For a drift space the apochromatic Twiss parameters
do not exist.}. 
In this paper we continue the study of the 
apochromatic Twiss parameters of straight drift-quadrupole systems
with special attention given to the properties of these parameters
for beamlines with symmetries.  

\vspace{-0.2cm}
\section{TRANSFER MAPS AND APOCHROMATIC TWISS PARAMETERS}

\vspace{-0.1cm}
Because we are interested in the lowest order
chromatic effects and because the map of the drift-quadrupole
system does not have second order geometric aberrations and
transverse coupling terms, we will
restrict our further consideration to the motion in one degree of freedom,
lets say, horizontal.
As usual, we will take the longitudinal particle position $\,\tau\,$ 
to be the independent variable and will use the variables
$\mbox{\boldmath $z$} = (x, p_x)^{\top}$
for the description of the horizontal beam oscillations.
Here $\,x\,$ is the horizontal coordinate and
$\,p_x\,$ is the horizontal canonical momentum scaled with 
the constant kinetic momentum of the reference particle.
To take into account energy dependence we will use
the variable $\,\varepsilon$, which is proportional to the relative
energy deviation, and we will treat this variable as a parameter.
With the assumptions made and with the precision needed the horizontal map
${\cal M}$ of the drift-quadrupole system can be represented through a Lie
factorization as follows 

\vspace{-0.1cm}
\noindent
\begin{eqnarray}
:{\cal M}: \,=_2\,
\exp(:- (\varepsilon \,/\, 2 ) \cdot {\cal Q}(x, \,p_x):) :M:,
\label{TWO_C_1}
\end{eqnarray}

\noindent
where $M$ is $2 \times 2$ linear transfer matrix,
the symbol $=_m$ denotes equality up to order $m$ (inclusive)
with respect to the variables $x$, $p_x$ and $\varepsilon$
when maps on both sides of (\ref{TWO_C_1})
are applied to the phase space vector $\mbox{\boldmath $z$}$,
and

\vspace{-0.2cm}
\noindent
\begin{eqnarray}
{\cal Q}(x, p_x) \,=\,
c_{20} \,x^2 \,+\, 2 c_{11} \,x p_x \,+\, c_{02} \,p_x^2.
\label{MKJ_2}
\end{eqnarray}

\vspace{-0.2cm}
\noindent
As it was shown in~\cite{ApIPAC10}, 
for every drift-quadrupole system which is not a pure drift space
the quadratic form ${\cal Q}$ is negative-definite and
the apochromatic Twiss parameters at the system entrance can be 
calculated using the coefficients of this quadratic form according to the following formulas

\vspace{-0.2cm}
\noindent
\begin{eqnarray}
\beta_a = -\frac{c_{02}}{\sqrt{c_{20} c_{02} -c_{11}^2}},
\;\;
\alpha_a = -\frac{c_{11}}{\sqrt{c_{20} c_{02} -c_{11}^2}}.
\label{MKJ_8}
\end{eqnarray}

\vspace{-0.4cm}
\section{CHROMATIC LATTICE FUNCTIONS}

\vspace{-0.1cm}
The equations of particle motion in a straight drift-quadrupole channel, 
when linearized with respect to the transverse variables, are uncoupled between 
the horizontal and vertical degrees of freedom, and 
we again can limit our considerations to the energy dependent
dynamics in one transverse dimension.
The $2 \times 2$ horizontal fundamental matrix $M_{\varepsilon}$ can be expressed 
with the help of the energy dependent lattice functions
$\beta_{\varepsilon}$, $\alpha_{\varepsilon}$, $\gamma_{\varepsilon}$
and $\mu_{\varepsilon}$,
if they are known, in the familiar form

\vspace{-0.2cm}
\noindent
\begin{eqnarray}
M_{\varepsilon}(l) \,=\,
T_{\varepsilon}^{-1}(l) \cdot R(\mu_{\varepsilon}(l)) \cdot T_{\varepsilon}(0),
\label{Equ_04}
\end{eqnarray}

\vspace{-0.2cm}
\noindent
where $\,R(\mu_{\varepsilon}(l))\,$ is a 2 by 2 rotation matrix and 

\vspace{-0.2cm}
\noindent
\begin{eqnarray}
T_{\varepsilon}(\tau) \,=\, 
\left(
\begin{array}{cc}
  1        / \sqrt{\beta_{\varepsilon}(\tau)} & 0 \\
\alpha_{\varepsilon}(\tau) / \sqrt{\beta_{\varepsilon}(\tau)} & 
\sqrt{\beta_{\varepsilon}(\tau)}
\end{array}
\right).
\label{Equ_05}
\end{eqnarray}

\vspace{-0.1cm}
\noindent
The parametrization (\ref{Equ_04}) is very widely used
in accelerator physics, because the plot of the Twiss parameters along
the beam line shows in the clear visual form the dynamics of the beam envelopes
or, depending on interpretation, the dynamics of the second order moments
of the beam distribution. The energy dependent lattice functions can be
either calculated numerically for any given value of the energy offset
or one can find them perturbatively using Taylor expansion
with respect to the variable $\varepsilon$.
The quantities which are typically used for the description 
of the first-order chromatic effects are chromaticity,
which for the drift-quadrupole system of the length $l$ can be 
determined as the integral

\vspace{-0.2cm}
\noindent
\begin{eqnarray}
\xi(\beta_0)
\,=\, 
- \frac{1}{2} \int_0^{l} \gamma_0 (\tau) \, d \tau,
\label{Equ_06}
\end{eqnarray}

\vspace{-0.2cm}
\noindent
and two betatron amplitude difference functions

\vspace{-0.15cm}
\noindent
\begin{eqnarray}
\hat{b}(\beta_{\varepsilon}) \,=\, 
\frac{1}{2} \cdot \left. \left( 
\frac{1}{\beta_{\varepsilon}}
\frac{d \beta_{\varepsilon}}{d \varepsilon}
\right) \right|_{\varepsilon = 0},
\label{Equ_07} 
\end{eqnarray}

\vspace{-0.3cm}
\noindent
\begin{eqnarray}
\hat{a}(\beta_{\varepsilon}) \,=\, 
\frac{1}{2} \cdot \left. \left( 
\frac{d \alpha_{\varepsilon}}{d \varepsilon} \,-\, 
\frac{\alpha_{\varepsilon}}{\beta_{\varepsilon}}
\frac{d \beta_{\varepsilon}}{d \varepsilon}
\right) \right|_{\varepsilon = 0},
\label{Equ_08} 
\end{eqnarray}

\noindent
which we denote with the hat on the top
in comparison with their usual notations ~\cite{MAD}
because for us it is convenient to scale them with the 
factor $1/2$. If the values of the chromatic variables $\hat{b}$, $\hat{a}$ and $\xi$
are known, then one can calculate the first (linear with respect to $\varepsilon$) 
correction terms for the Twiss parameters and for the phase advance 
one finds

\vspace{-0.1cm}
\noindent
\begin{eqnarray}
\left.  
\frac{d \mu_{\varepsilon}(l)}{d \varepsilon} \right|_{\varepsilon = 0}
\,=\,
\xi(\beta_0) \,+\, \hat{a}(\beta_{\varepsilon}(0)) \,-\, \hat{a}(\beta_{\varepsilon}(l)).
\label{Equ_09} 
\end{eqnarray}

\vspace{-0.1cm}
\noindent
One sees that three variables (\ref{Equ_06})-(\ref{Equ_08}) are,
in general, sufficient for the description of all first order chromatic effects.
Nevertheless, there is still a need for additional variables to make this description
more detailed. It is connected with the fact that the betatron amplitude difference functions
$\hat{b}$ and $\hat{a}$ play a twofold role. They describe both, chromatic properties
of the incoming Twiss parameters and interaction of the incoming Twiss parameters
with the chromatic properties of the given beam line. 
So in the paper ~\cite{ApIPAC10} we have introduced two additional (supplementing) 
chromatic variables $\eta$ and $\zeta$, which we call apochromaticities 
and which (together with the chromaticity $\xi$) can be obtained
as coefficients of the expansion of the quadratic form (\ref{MKJ_2})
with respect to three independent quadratic invariants of
linear motion. 
These variables satisfy the important equality

\vspace{-0.1cm}
\noindent
\begin{eqnarray}
\xi^2(\beta_0)\,=\,
\eta^2(\beta_0) \,+\,\zeta^2(\beta_0)  \,+\,
\left(c_{20}\, c_{02} \,-\, c_{11}^2 \right),
\label{Equ_10}
\end{eqnarray}

\vspace{-0.1cm}
\noindent
and with their help the change in the betatron amplitude
difference functions
after the beam line passage
can be calculated as follows 

\vspace{-0.1cm}
\noindent
\begin{eqnarray}
\hat{b}(\beta_{\varepsilon}(l)) 
\,+\, 
i \,\hat{a}(\beta_{\varepsilon}(l)) \,=\,
\exp(i  2 \mu_0(l)) \cdot
\nonumber
\end{eqnarray}

\vspace{-0.2cm}
\noindent
\begin{eqnarray}
\Bigl[
\Bigl(
\hat{b}(\beta_{\varepsilon}(0)) \,+\,i\,
\hat{a}(\beta_{\varepsilon}(0))
\Bigr) 
\,+\, 
\Bigl(
\eta(\beta_0) \,+\,i\, \zeta(\beta_0)
\Bigr) 
\Bigr].
\label{CH_03} 
\end{eqnarray}

\vspace{-0.1cm}
\noindent
Taking into account the formula for the 
mismatch between 
the on and the off energy Twiss parameters

\vspace{-0.2cm}
\noindent
\begin{eqnarray}
m_p\left(\beta_{\varepsilon},\, \beta_0 \right)
\,=\,
\left(\beta_{\varepsilon}\, \gamma_0 \,-\, 2\, \alpha_{\varepsilon}\, \alpha_0 
\,+\, \gamma_{\varepsilon}\,\beta_0 \right) \,/\, 2
\,=
\nonumber
\end{eqnarray}

\vspace{-0.2cm}
\noindent
\begin{eqnarray}
1 \,+\, 2 \,\varepsilon^2 \cdot 
\left(\hat{a}^2(\beta_{\varepsilon}) \,+\, 
\hat{b}^2(\beta_{\varepsilon}) \right) \,+\, O(\varepsilon^3),
\label{CH_02} 
\end{eqnarray}

\vspace{-0.1cm}
\noindent
one sees from (\ref{CH_03}) that if on the system
entrance the incoming Twiss parameters for the particles with the nominal energy
were apochromatic with
$\,\eta(\beta_0) = \zeta(\beta_0) =0$, then 
the mismatch at the beam line exit will be the same as at the beam line
entrance (at least in the lowest order with respect to the energy deviation),
which (once more) shows the importance of the apochromatic Twiss parameters
for practical accelerator designs. 

To finish this section, let us note that if the apochromatic Twiss parameters
$\beta_a$ and $\alpha_a$ and the chromaticity $\xi(\beta_a)$
are known, then for an arbitrary Twiss parameters 
$\beta_0$ and $\alpha_0$ the following relations hold

\vspace{-0.1cm}
\noindent
\begin{eqnarray}
\xi(\beta_0) \,=\, \xi(\beta_a) \cdot m_p,
\label{CH_04} 
\end{eqnarray}

\vspace{-0.3cm}
\noindent
\begin{eqnarray}
\eta(\beta_0) \,=\, -\xi(\beta_a) \cdot \sqrt{m_p^2 - 1} \cdot \sin(2\theta).
\label{CH_05} 
\end{eqnarray}

\noindent
\begin{eqnarray}
\zeta(\beta_0) \,=\, -\xi(\beta_a) \cdot \sqrt{m_p^2 - 1} \cdot \cos(2\theta),
\label{CH_06} 
\end{eqnarray}

\vspace{-0.1cm}
\noindent
Here $m_p = m_p(\beta_0,\, \beta_a)$ and $\theta$ is the mismatch phase with

\vspace{-0.2cm}
\noindent
\begin{eqnarray}
\sin(2\theta) = \frac{1}{\sqrt{m_p^2 - 1}} \cdot
\left(\alpha_0(0)\, \frac{\beta_a(0)}{\beta_0(0)} \,-\, \alpha_a(0) \right),
\label{CH_07} 
\end{eqnarray}

\vspace{-0.3cm}
\noindent
\begin{eqnarray}
\cos(2\theta) = \frac{1}{\sqrt{m_p^2 - 1}} \cdot
\left(\frac{\beta_a(0)}{\beta_0(0)} \,-\, m_p  \right).
\label{CH_08} 
\end{eqnarray}

\vspace{-0.3cm}
\section{PERIODIC SYSTEMS}

\vspace{-0.1cm}
In this section we will consider a system constructed by a repetition of
$n$ identical cells ($n > 1$) with the assumption that the
linear (energy independent) cell transfer matrix allows periodic beam transport with
the periodic  Twiss parameters $\beta_0^p$
and $\alpha_0^p$, and with the periodic cell phase advance $\mu_0^p$ which is 
not a multiple of $\pi$.
Let $\eta(\beta_0^p)$ and  $\zeta(\beta_0^p)$
be the one cell apochromaticities.  
Iterating the propagation formula (\ref{CH_03}) 
$n$ times, we obtain

\vspace{-0.1cm}
\noindent
\begin{eqnarray}
\hat{b}(\beta_{\varepsilon}^p(nl)) \,+\, i \,\hat{a}(\beta_{\varepsilon}^p(nl)) \,=
\nonumber
\end{eqnarray}

\vspace{-0.3cm}
\noindent
\begin{eqnarray}
\exp(i 2 n \mu_0^p) \cdot
\Bigl[
\Bigl(
\hat{b}(\beta_{\varepsilon}^p(0)) \,+\, 
i \,\hat{a}(\beta_{\varepsilon}^p(0))
\Bigr)
\,+
\nonumber
\end{eqnarray}

\vspace{-0.3cm}
\noindent
\begin{eqnarray}
\frac{\sin(n \mu_0^p)}{\sin(\mu_0^p)}
\cdot 
\exp(-i (n-1) \mu_0^p) 
\cdot
\Bigl(
\eta(\beta_0^p) + i\, \zeta(\beta_0^p)
\Bigr)
\Bigr].
\label{PS_01} 
\end{eqnarray}

\vspace{-0.1cm}
\noindent
The second addend inside the square brackets in the right hand side
of this formula
gives us the $n$-cell apochromaticities 
and one sees that if the $n$-cell phase advance $n \mu_0^p$
is multiple of $\pi$, then 
these apochromaticities are equal to zero.
So we came to the following statement, which is valid for each
transverse plane separately:

For any system built out of $n$ identical drift-quadrupole cells 
with the overall transfer matrix equal to the identity
or to the minus identity matrix and with the cell matrix not
equal to the identity
or to the minus identity matrix, the cell periodic
Twiss parameters (which are unique under assumptions made)
are unique apochromatic Twiss parameters of the $n$-cell system.

Note that, though not in such general form and without addressing the question of uniqueness,
this statement cannot be considered as a completely new
result. For example in ~\cite{Zotter} the same was shown for the sequence
of FODO cells by making explicit calculations involving thin-lens model for the
quadrupole focusing.

\vspace{-0.3cm}
\section{SCALING AND TELESCOPES}

\vspace{-0.1cm}
Scaling of lattice parameters is a procedure which allows
to adapt known optics solutions to the new geometrical dimensions.
For a drift-quadrupole system scaling by a factor $\lambda > 0$ consists of 
the elongation of the system length by the factor $\lambda$ with
the simultaneous reduction of the quadrupole coefficient $k_1(\tau)$
by the factor $\lambda^2$. The usefulness of this procedure is
determined by the fact that any betatron function of the original system,
when elongated to the length of the new system and multiplied by
the factor $\lambda$, becomes the betatron function of the scaled system.

The map of the scaled system 

\vspace{-0.1cm}
\noindent
\begin{eqnarray}
:{\cal M}_{\lambda}: \,=_2\,
\exp(:- (\varepsilon \,/\, 2 ) \cdot {\cal Q}_{\lambda}(x, \,p_x):) :M_{\lambda}:,
\label{ST_1}
\end{eqnarray}

\vspace{-0.1cm}
\noindent
is connected to the map of the original system (\ref{TWO_C_1})
by the relation

\vspace{-0.1cm}
\noindent
\begin{eqnarray}
:{\cal M}_{\lambda}: \,=_2\,
:\Lambda(\lambda):^{-1} :{\cal M}: :\Lambda(\lambda):,
\label{ST_2}
\end{eqnarray}

\vspace{-0.1cm}
\noindent
where 
$\Lambda(\lambda) \,=\,\mbox{diag}(\sqrt{\lambda}, \,1/\sqrt{\lambda})\,$
is the scaling matrix.
In particular, the quadratic form ${\cal Q}_{\lambda}$ is given by the
formula

\vspace{-0.1cm}
\noindent
\begin{eqnarray}
{\cal Q}_{\lambda}(x, p_x) =
\lambda^{-1} c_{20} \,x^2 \,+\, 2 c_{11} \,x p_x \,+\, \lambda\,c_{02} \,p_x^2, 
\label{ST_4}
\end{eqnarray}

\vspace{-0.1cm}
\noindent
and, 
applying formulas (\ref{MKJ_8}) to the coefficients
of this quadratic form,
one sees that the apochromatic Twiss parameters
have the same scaling properties as any other betatron functions,
as could be expected.

The scaling is not only important by itself, it also gives a 
systematic way for the design of beam magnification (or demagnification)
telescopes.
Let us consider a system constructed by a repetition of
$n$ cells and let us assume that the cell with the index $m$ ($m = 2, \ldots, n$) is
the copy of the first cell scaled by the factor $\lambda^{m-1}$.
The overall transfer matrix of this system is equal to
the matrix 

\vspace{-0.1cm}
\noindent
\begin{eqnarray}
\Lambda^{n}(\lambda) \cdot (\Lambda^{-1}(\lambda) \, M)^n,
\label{ST_5}
\end{eqnarray}

\vspace{-0.1cm}
\noindent
and, if the second multiplier in the product (\ref{ST_5}) is equal to
the plus or minus identity matrix, then the system becomes a telescope.

So let us assume that

\vspace{-0.1cm}
\noindent
\begin{eqnarray}
(\Lambda^{-1}(\lambda) \, M)^n \,=\, \delta \cdot I,
\;\;\;\;\;\;
\delta\,=\,\pm 1.
\label{ST_5_1}
\end{eqnarray}

\vspace{-0.1cm}
\noindent
Then the map of the total system $\,{\cal M}_{t}\,$ takes the form

\vspace{-0.1cm}
\noindent
\begin{eqnarray}
:{\cal M}_{t}: \,=_2\,
\exp(:- (\varepsilon \,/\, 2 ) \cdot {\cal S}_{\lambda}(x, \,p_x):)
:\delta \,\Lambda^n(\lambda):,
\label{ST_6}
\end{eqnarray}

\vspace{-0.1cm}
\noindent
where the quadratic form 

\vspace{-0.1cm}
\noindent
\begin{eqnarray}
{\cal S}_{\lambda}(x, \,p_x) \,=\,
\sum\limits_{m = 0}^{n-1}
:\Lambda^{-1}(\lambda) \, M:^m
{\cal Q}(x, \,p_x)
\label{ST_7}
\end{eqnarray}

\vspace{-0.1cm}
\noindent
is negative-definite (as for any drift-quadrupole system)
and, clearly, is the invariant of the cyclic group generated by the matrix
$\,\Lambda^{-1}(\lambda) \, M$. 
If this cyclic group has more than two elements, then 
$\,{\cal S}_{\lambda}\,$ must be proportional to the 
Courant-Snyder quadratic form which corresponds to the (unique in this case)
periodic Twiss parameters of the matrix $\,\Lambda^{-1}(\lambda) \, M$. 
Thus we have proven the following:

If the condition (\ref{ST_5_1}) is satisfied and if the matrix
$\,\Lambda^{-1}(\lambda) \, M\,$ is not equal to the identity or
to the minus identity matrix, then the apochromatic Twiss parameters
of the considered $n$-cell telescope coincide with the periodic
Twiss parameters of the matrix $\,\Lambda^{-1}(\lambda) \, M$.

Note that for $\lambda = 1$ this result recovers the statement of
the previous section concerning $n$-cell periodic systems.
Note also that apochromatic properties of some
telescopes constructed by scaling 
on the basis of the doublet cell
were studied in ~\cite{MontRugg}
by using thin-lens approximation,
and the possibility to make the scaled $n$-cell system a second
order achromat by introducing bending of the central trajectory
and sextupole fields was addressed in ~\cite{BrownServranckx}.

\section{MIRROR SYMMETRY}

\vspace{-0.1cm}
Let us consider a mirror symmetric system of the length $2 l$
where the drift-quadrupole cell which is not a pure drift space is
followed by its reversed image and let us assume that the map
of the first part is given by the expression (\ref{TWO_C_1}). 
Then the map of the second (reversed) part can be calculated as follows

\vspace{-0.1cm}
\noindent
\begin{eqnarray}
:{\cal M}_{r}: \,=_2\,
\exp(:- (\varepsilon \,/\, 2 ) \cdot {\cal Q}_r(\mbox{\boldmath $z$}):) 
:T_r M^{-1} T_r:,
\label{MS_1}
\end{eqnarray}

\vspace{-0.1cm}
\noindent
where 
$\,{\cal Q}_r(\mbox{\boldmath $z$}) = {\cal Q}(M^{-1} T_r \,\mbox{\boldmath $z$})\,$
and
$\,T_r = \mbox{diag}(1, -1)\,$
is the reversal symmetry matrix.
And for the map of the total system one finds

\vspace{-0.1cm}
\noindent
\begin{eqnarray}
:{\cal M}_{t}: =_2
\exp(:- (\varepsilon / 2 ) \cdot {\cal Q}_t(\mbox{\boldmath $z$}):) 
:T_r M^{-1} T_r M:,
\label{MS_2}
\end{eqnarray}

\vspace{-0.1cm}
\noindent
where

\vspace{-0.1cm}
\noindent
\begin{eqnarray}
{\cal Q}_t(\mbox{\boldmath $z$}) \,=\,
{\cal Q}(\mbox{\boldmath $z$})
\,+\,
{\cal Q}(M^{-1} T_r M \,\mbox{\boldmath $z$}).
\label{MS_3}
\end{eqnarray}

\vspace{-0.1cm}
\noindent
The quadratic form ${\cal Q}_t$ uniquely determines the
apochromatic Twiss parameters of the total system at the
system entrance and, instead of the study of the evolution
of the apochromatic Twiss parameters along the beamline,
one can equivalently analyze the propagation of this 
quadratic form along the system length.
Let us look at the images of the quadratic form ${\cal Q}_t$ 
at the system symmetry point (in the middle of the system)

\vspace{-0.1cm}
\noindent
\begin{eqnarray}
:M:^{-1}\,{\cal Q}_t(\mbox{\boldmath $z$}) \,=\,
{\cal Q}(M^{-1}\,\mbox{\boldmath $z$}) \,+\,
{\cal Q}(M^{-1} T_r\,\mbox{\boldmath $z$})
\label{MS_4}
\end{eqnarray}

\vspace{-0.1cm}
\noindent
and at the system exit

\vspace{-0.1cm}
\noindent
\begin{eqnarray}
:T_r M^{-1} T_r M:^{-1}\,{\cal Q}_t(\mbox{\boldmath $z$}) \,=\,
{\cal Q}_t(T_r\,\mbox{\boldmath $z$}).
\label{MS_5}
\end{eqnarray}

\vspace{-0.1cm}
\noindent
The formulas (\ref{MS_4}) and (\ref{MS_5}) tell us that
the apochromatic Twiss parameters of the mirror symmetric
system are mirror symmetric by itself, i.e.

\vspace{-0.1cm}
\noindent
\begin{eqnarray}
\beta_a(2 l) = \beta_a(0),
\;\;
\alpha_a(2 l) = -\alpha_a(0),
\;\;
\alpha_a(l) = 0,
\label{MS_6}
\end{eqnarray}

\vspace{-0.1cm}
\noindent
as (like for the scaling) also could be expected.

Let us finish this paper with the remark that
if one will consider the both parts of the system
separately, then one will find that the entrance apochromatic Twiss
parameters of the second (reversed) part are the reversed image of the
exit apochromatic Twiss parameters of the first part,
and that the apochromatic Twiss parameters of the total system will
coincide with the apochromatic Twiss parameters of both parts
if, and only if, the exit apochromatic alpha of the first part 
is equal to zero.

\end{document}